\pdfoutput=1
\documentclass[doublespacing]{elsart}

\usepackage{graphicx}
\usepackage{amssymb}
\usepackage{amsmath}
\usepackage{lineno}
\usepackage{epsfig}  
\usepackage{color}
\usepackage{graphicx}
\usepackage{graphics} 
\usepackage{dcolumn} 
\usepackage{bm}
\usepackage{amsfonts}
\begin{document}

\begin{frontmatter}

\title{Tuning Mixing within a Droplet for Digital Microfluidics}
\author[RC]{R. Chabreyrie}
\author[DV1,DV2]{D. Vainchtein}
\author[CC]{C. Chandre}
\author[PS]{P. Singh}
\author[RC]{N. Aubry}
\address[RC]{Mechanical Engineering Department, Carnegie Mellon University, Pittsburgh, PA 15213, USA}
\address[DV1]{School of Physics, Georgia Institute of Technology, GA 30332, USA}
\address[DV2]{Space Research Institute, Moscow, GSP-7, 117997, Russia}
\address[CC]{Centre de Physique Th\'eorique\thanksref{CNRS}, CNRS -- Aix-Marseille Universit\'es, Luminy-case 907, F-13288 Marseille cedex 09, France}
\address[PS]{Mechanical Engineering Department, New Jersey Institute of Technology, Newark, NJ 07102, USA}
\thanks[CNRS]{UMR 6207 of the CNRS, Aix-Marseille and Sud Toulon-Var
Universities. Affiliated with the CNRS Research Federation FRUMAM (FR 2291).
CEA registered research laboratory LRC DSM-06-35.}

\begin{abstract}
The design of strategies to generate efficient mixing is crucial for a
variety of applications, particularly digital microfluidic devices that use
small ``discrete'' fluid volumes (droplets) as fluid carriers and microreactors.
In recent work, we have presented an approach for the generation and control of
mixing inside a translating spherical droplet. This was accomplished by considering
Stokes' flow within a droplet proceeding downstream to which we have superimposed time
dependent (sinusoidal) rotation.  The mixing obtained is the result of the stretching and
folding of material lines which increase exponentially the surface contact between
reagents. The mixing strategy relies on the generation of resonances between the steady
and the unsteady part of the flow, which is achieved by tuning the parameters of the
periodic rotation.  Such resonances, in our system, offer the possibility of controlling both the
location and the size of the mixing region within the droplet, which may be useful to manufacture
inhomogeneous particles (such as Janus particles). While the period and amplitude of the
periodic rotation play a major role,  it is shown here by using a triangular function that the particular
shape of the rotation (as a function of time) has  a
minor influence. This finding demonstrates the robustness of the proposed mixing strategy, a crucial
point for its experimental realization.
\end{abstract}

\end{frontmatter}

\section{Introduction}
Although most microfluidics have been using fluid-streams as the main means to carry fluids, devices based on individual droplets have been proposed as well. In the latter, also called ``digital microfluidic" systems, ``discrete'' fluid volumes
(droplets) rather than continuous streams are used, with the potential
to utilize individual droplets as microreactors \cite{Ismagilov:2003}.\\
Whether fluids are encapsulated within droplets or flow along channels, reactions can occur efficiently only in presence of rapid mixing. Such mixing conditions are not easy to fulfill due to the low
Reynolds number of the flow involved, which prevents turbulence from taking place. Stirring, in addition to molecular diffusion, is needed to stretch and fold fluid elements, thus significantly increasing interfacial areas.  Whereas some strategies are based on complex channel geometry for flows in microchannels, active methods (via external forcing) (see, e.g. \cite{Oddy:2001,Bau:2001,Ouldelmoctar:2003,GlasgowAubry:2003,Glasgow:2004}) have
also resulted in efficient mixing, especially at very low Reynolds numbers
\cite{Goullet:2006}. The combination of both geometry alteration and forcing has been explored
as well \cite{Goullet:2006,Niu:2003,Bottausci:2004,Stremler:2004,Lee:2007}.\\
Chaotic advection inside a liquid droplet subjected to a forcing has been studied extensively
\cite{Baj,KroujilineandStone:1999,Lee:2000,WardandHomsy:2001,Grigoriev:2005,
Homsy:2007,VWG:2007} and obtained experimentally by means of oscillatory flows
\cite{WardandHomsy:2003,GSS:2006}. In this paper as in \cite{ChabreyriePRE08}, we concentrate on controlling both the size and the location of the mixing. Indeed, we are interested in both complete and incomplete mixing within drops.   While complete mixing should be useful for efficient reactions to occur uniformly, there are applications in which non-uniform mixing, when well-controlled, could be desired.   The latter include the formation of inhomogeneous drops and particles, such as Janus particles, whose properties have been shown, in certain cases, to be superior to their homogeneous counterparts.  In this paper, we extend the results
of  \cite{ChabreyriePRE08} which were obtained  by using unsteady -- sinusoidal --
forcing on a translating droplet.  Specifically, we study the robustness of the control of the size and location of the mixing region by selecting a different time function for the rotation.\\
The existence of chaotic behavior in three-dimensional bounded steady flows has
been shown (e.g. \cite{Baj,KroujilineandStone:1999,VVN:1996a,VNM:2006}). Our
work is distinct from the latter contributions through the addition of unsteadiness,
which is crucial to control the chaotic mixing behavior through resonance effects \cite{Lima:1990,CFP2:1996,VWG:2007}.\\
The physical system under consideration in this work is described in
Sec.~\ref{Pys. Sys.}, and our numerical results are displayed in
Sec.~\ref{results}. The control of both the location (Sec.~\ref{location}) and size
(Sec.~\ref{size}) of the chaotic mixing region is studied qualitatively via
{\it Liouvillian} sections.
\section{ Physical system \label{Pys. Sys.}}
\subsection{Flow equations and assumptions}
Let us assume a Newtonian liquid droplet of sperical shape suspended in an
incompressible Newtonian fluid. Its motion consists of a translation and a slow
rigid body rotation (see \cite{KroujilineandStone:1999}). As in the
previous reference, we assume that the interfacial tension is sufficiently
large for the drop to remain spherical through its motion and that the Reynolds number is very small
compared to $1$. Thus, a reasonable approximation is to consider that
both the internal and external flows are Stokes flows. The boundary conditions at the
droplet surface can be derived from the continuity of velocity and tangential stress conditions.\\
The resulting internal flow is a superimposition of a steady base flow
(non-mixing flow) and an unsteady rigid-body rotation. Consider a Cartesian
coordinate system translating with the center-of-mass velocity of the droplet
with the orientation of the axes such that the unit vector $\bm{e}_{z}$ points in the
direction of the translation and the unit vector $\bm{e}_{x}$ lies in the
$\bm{\omega}-\bm{e}_{z}$ plane. We then obtain the following unsteady internal
flow:
\begin{eqnarray}
\label{veloT21}
u&=& zx - \epsilon a_{\omega}(t) \omega_z y,\\
\label{veloT22}
v&=& zy + \epsilon a_{\omega}(t)\left(\omega_z x - \omega_x z\right),\\
\label{veloT23}
w&=& 1-2x^2-2y^2-z^2 + \epsilon a_{\omega}(t)\omega_x y,
\end{eqnarray}
where $u$, $v$ and $w$ are the time derivatives of $x$, $y$ and $z$, respectively.
In Eqs.~(\ref{veloT21}-\ref{veloT23}), all the lengths and velocities were made
dimensionless by using the droplet radius and the magnitude of the
translational velocity as the length and velocity scales, respectively. The
rotation is characterized by a rotation with a maximum amplitude $\epsilon \ll
1$, a fixed orientation vector
\begin{equation}
\bm{\omega}=\left(\sqrt{2}/2,0,\sqrt{2}/2\right),
\nonumber
\end{equation}
and a periodic amplitude of the rotation $a_{\omega}(t)$. In the present paper
we study two time dependent functions for the amplitude of the rotation $a_{\omega}(t)$, one
consisting of only one harmonic:
\begin{equation}
\label{cos}
{\mathcal C}_{\omega}(t)=\frac{1}{2}\left(1+\cos \omega t \right),
\end{equation}
and the other one consisting of a triangular function with an infinite
 number of harmonics:
\begin{equation}
\label{tri}
{\mathcal T}_{\omega}(t) =
1+\sum^{\infty}_{n=1}\alpha_{2n-1}\cos\left(\left(2n-1\right)\omega t\right),
~~\mbox{where}~~\alpha_n=\left(\frac{2}{n\pi}\right)^{2}.
\end{equation}
Equations (\ref{veloT21}-\ref{veloT23}) are the same as in
\cite{KroujilineandStone:1999} except that the vorticity vector, that was
constant in \cite{KroujilineandStone:1999}, is  unsteady in our case.
Note that
the internal flow is a solution of the (time dependent) Stokes problem in which
a time-dependent body force has been added on the right hand side of the momentum equation.  In practice, this time-dependent forcing could be realized,
e.g. by creating a time-dependent swirl motion in the external flow or by
applying an electric field that exerts a torque on the drop (see,
e.g. \cite{AubrySingh:2006} or work on electrorotation \cite{Arn:1988}).
Note that the surface of the droplet, $r^2 = x^2+y^2+z^2 =1$ is invariant under flow
(\ref{veloT21}-\ref{veloT23}).
\subsection{Non-mixing case}
\begin{figure}
\begin{center}
\includegraphics*[width=14cm]{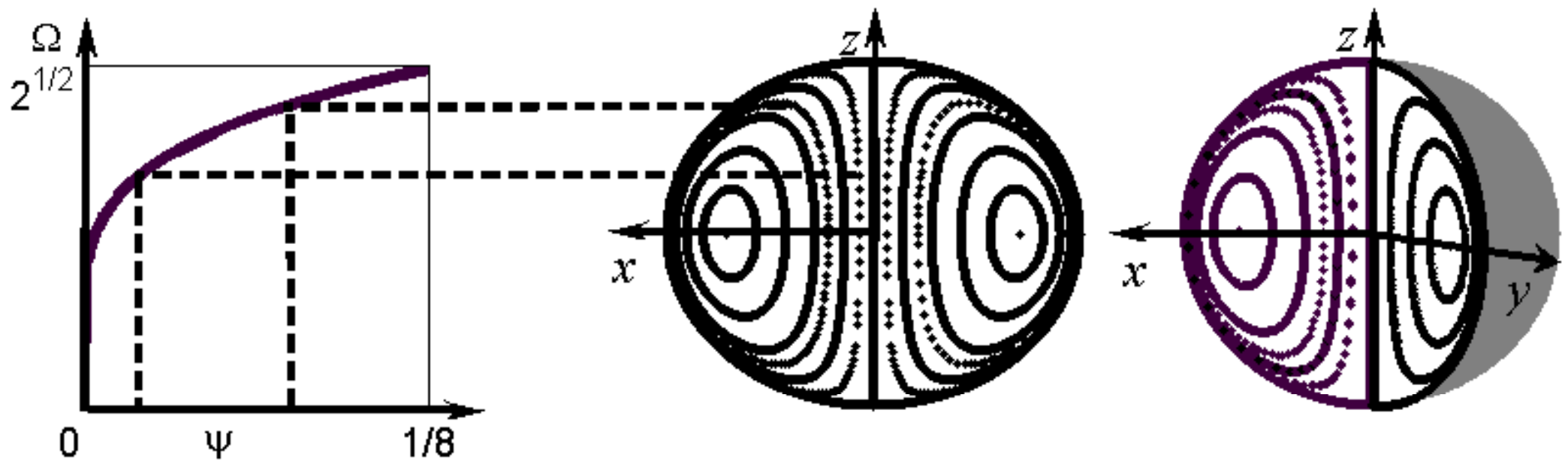}
\end{center}
\caption{\label{figure1} Streamlines in a cross-section of the drop (without
rotation) and their frequencies $\Omega\left(\psi\right)$ as given by
Eq.~(\ref{Per}).}
\end{figure}

The non-mixing (unperturbed) case or base flow, i.e. $\epsilon =0$, is characterized by two
invariants (i.e. time-independent quantity), the stream function $\psi$ and the
azimuthal angle $\phi$ such that:
\begin{equation}
\psi = \frac{1}{2} \left(x^2+y^2 \right)\left(1-x^2-y^2-z^2\right), \quad \phi =\arctan \left(y/x\right),
\nonumber
\end{equation}
where $\psi \in \left[0,1/8\right]$ and $\phi \in [0,2\pi[$. The streamlines of
the non-mixing case are joint lines of constant $\psi$ and $\phi$, denoted by
$\Gamma_{\psi,\phi}$. The heteroclinic orbits $\Gamma_{\psi,\phi}$ such that $\psi=0$ and $\phi \in [0,2\pi[$,
connect two hyperbolic fixed points located at the poles of the sphere (see
Fig.~\ref{figure1}). All other streamlines are closed curves that converge
toward a circle of degenerate elliptic fixed point $(x^2+y^2=1/2, z=0)$ as
$\psi$ is increased toward the value $1/8$. The frequency of motion on the
streamline $\Gamma_{\psi,\phi}$ is independent of $\phi$ and given by
\begin{equation}
\label{Per}
\frac{2\pi}{\Omega(\psi)} = \int^{\pi/2}_{-\pi/2}\frac{\sqrt{2} \;
d \beta}{\sqrt{1+\gamma(\psi)\sin \beta}} =\frac{2\sqrt{2}}{\sqrt{1+\gamma}}
K\left( \sqrt{\frac{2\gamma}{1+\gamma}}\right),
\end{equation}
with $\gamma(\psi)=\sqrt{1-8\psi}$ and $K$ is the complete elliptic function
of the first kind. The plot of $\Omega$ as a function of $\psi$, presented in
Fig.~\ref{figure1}, shows that $\Omega$ is bounded by two limits,
$\Omega(0)=0$ and $\Omega(1/8)=\sqrt{2}$. On every streamline
$\Gamma_{\psi,\phi}$ we introduce a uniform phase $\chi  \bmod 2\pi$
such that $\chi=0$ on the ${\bm e}_{x}-{\bm e}_{y}$ plane and
$\dot{\chi}=\Omega\left(\psi\right)$. With this uniform phase, the non-mixing
flow can be described by using the variables $(\psi,\phi,\chi)$ instead of the
Cartesian coordinates $(x,y,z)$: \[\dot{\psi} =0, \quad \dot{\phi} =0, \quad
\dot{\chi} = \Omega(\psi).\] Such a system is generally classified as {\it
action-action-angle}, where the two invariants $\psi$ and $\phi$ are the two
actions and $\chi$ is the angle.
\subsection{Mixing case}
In the mixing (or weakly perturbed) case $0<\epsilon\ll 1$, the quantities
$\psi$ and $\phi$ are no longer constant and their time evolution satisfies
\begin{equation}
\dot{\psi}=-2 \epsilon a_{\omega}(t)\omega_x\psi\sin\phi G\left(\psi,\chi\right),
\nonumber
\end{equation}
\begin{equation}
\dot{\phi}=\epsilon a_{\omega}(t)\omega_z - \epsilon a_{\omega}(t)\omega_x\cos\phi G\left(\psi,\chi\right),
\nonumber
\end{equation}
where $G(\psi,\chi)=z/(x^2+y^2)$ is $2\pi$ periodic in $\chi$ and has zero
average in $\chi$. The time evolution equation for $\chi$ reads
\begin{equation}
\dot{\chi}=\Omega(\psi)+\epsilon a_{\omega}(t) H(\psi,\phi, \chi),
\nonumber
\end{equation}
where $H$ is $2\pi$ periodic in $\chi$. Such a system possesses two time
scales, a fast one (of order one) associated with $\chi$, typically on time
scales of $1$, and a slow one (of order $1/\epsilon$) associated with both
$\psi$ and $\phi$. The generation of chaos arises from the resonances between
the frequency of the integrable case $\Omega (\psi)$ and the forcing frequency
$\omega$.
\section{Generation and control of the chaotic mixing region \label{results} }
In this work we fallow the approach of  \cite{ChabreyriePRE08} that
enables the generation of a three-dimensional chaotic mixing region,
for which we are able to control both the location and the size. The method consists in
bringing a specific family of the unperturbed tori $ \left\{
\Gamma_{\psi_{n,m}}\right\}_{n,m~\in \mathbb{N}^{2}}$ into
resonance with the periodic perturbation $a_{\omega}(t)$ by
selecting the frequency $\omega$ so that it satisfies the resonance condition:
\begin{equation}
n\Omega\left(\psi_{n,m}\right)-m\omega=0,~~\mbox{for}~~n,m \in \mathbb{N}^2.
\nonumber
\end{equation}
The control is realized by selecting the two parameters that
characterize the periodic rigid body rotation, specifically its the maximum
amplitude $\epsilon$ and its frequency $\omega$.
The amplitude satisfies $0\leq\epsilon<<1 $, where
the lower and upper limits correspond to the absence of mixing and
complete mixing, respectively.  In this work we are interested in the location of the mixing
regions within the drop.  To study our system, we compute two-dimensional projections
of time-periodic three-dimensional flows via the  combination of a stroboscopic map
and a plane section (here, the $y=0$ plane). In other words, the points on these so called Liouivillian sections are
 the intersections of the trajectories with the plane $y=0$ at every period $2\pi/\omega$.

\begin{figure}
\begin{center}
\includegraphics*[width=14cm]{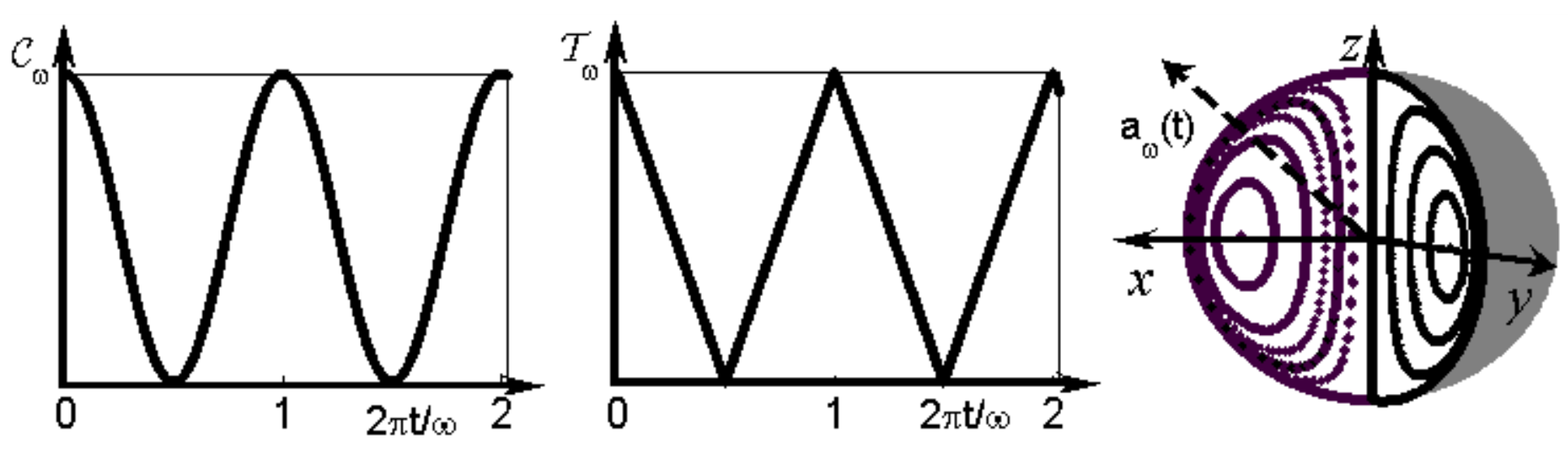}
\end{center}
\caption{\label{cos_tri}Right panel: Spherical droplet
represented with rigid body rotation (axis in dash line).
Middle panel: Plot of the periodic forcing ${\mathcal
T}_{\omega}(t)$ of the rigid body rotation.
Left panel: Plot of the periodic forcing ${\mathcal
C}_{\omega}(t)$ of the rigid body rotation.}
\end{figure}
\subsection{Control of the location \label{location}}
\begin{figure}
\begin{center}
\includegraphics*[width=14cm]{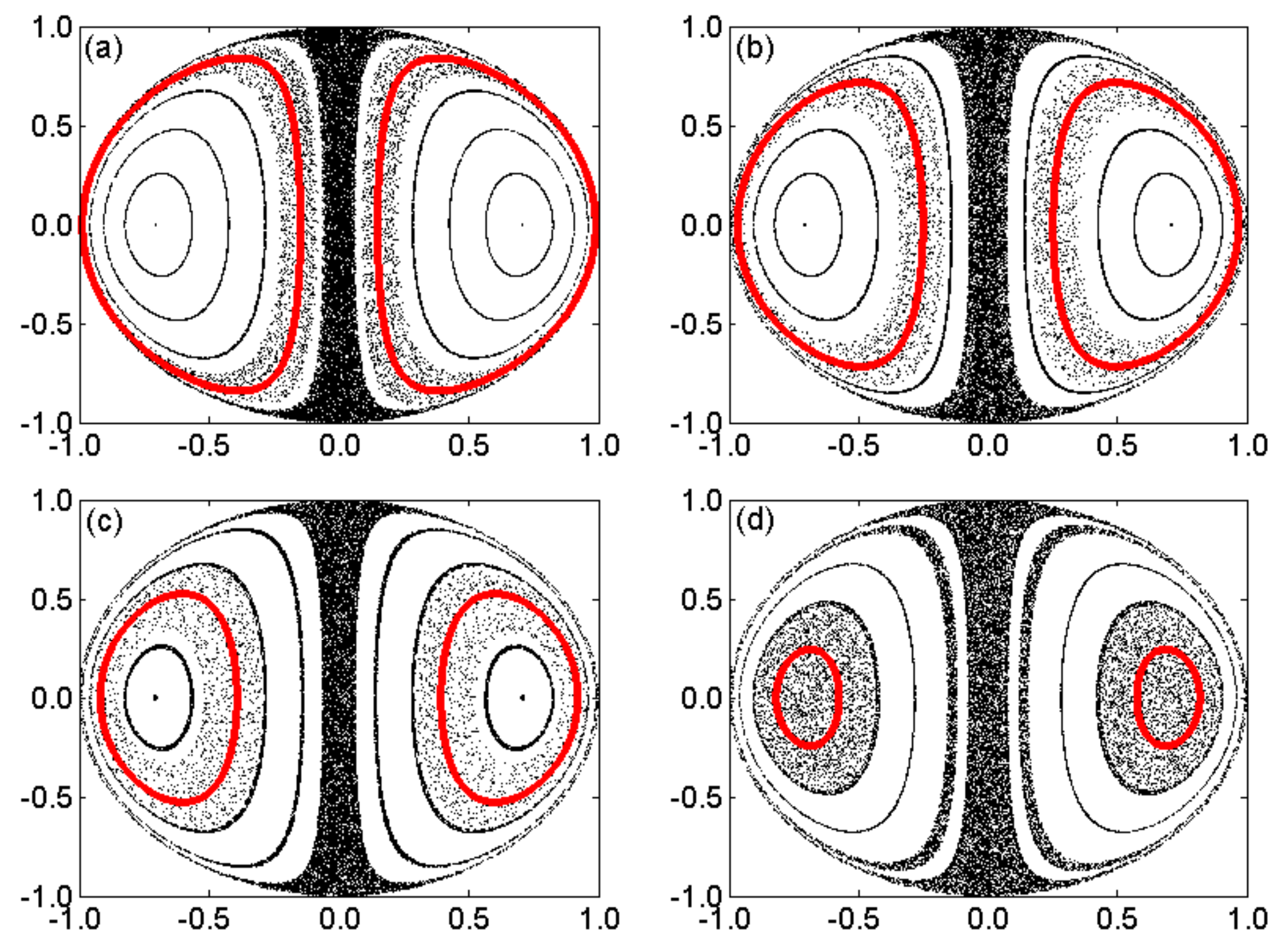}
\end{center}
\caption{\label{figure2} (Color on line)
Liouvillian sections for the time-dependence ${\mathcal C}_{\omega}$ with the
frequencies $\omega= 0.95, 1.10, 1.25, 1.40$ (a-d) and the amplitude $\epsilon
= 0.05$. The (red) dashed line inside the primary mixing region is
$\Gamma_{\psi_{1,1}}$.}
\end{figure}

\begin{figure}
\begin{center}
\includegraphics*[width=14cm]{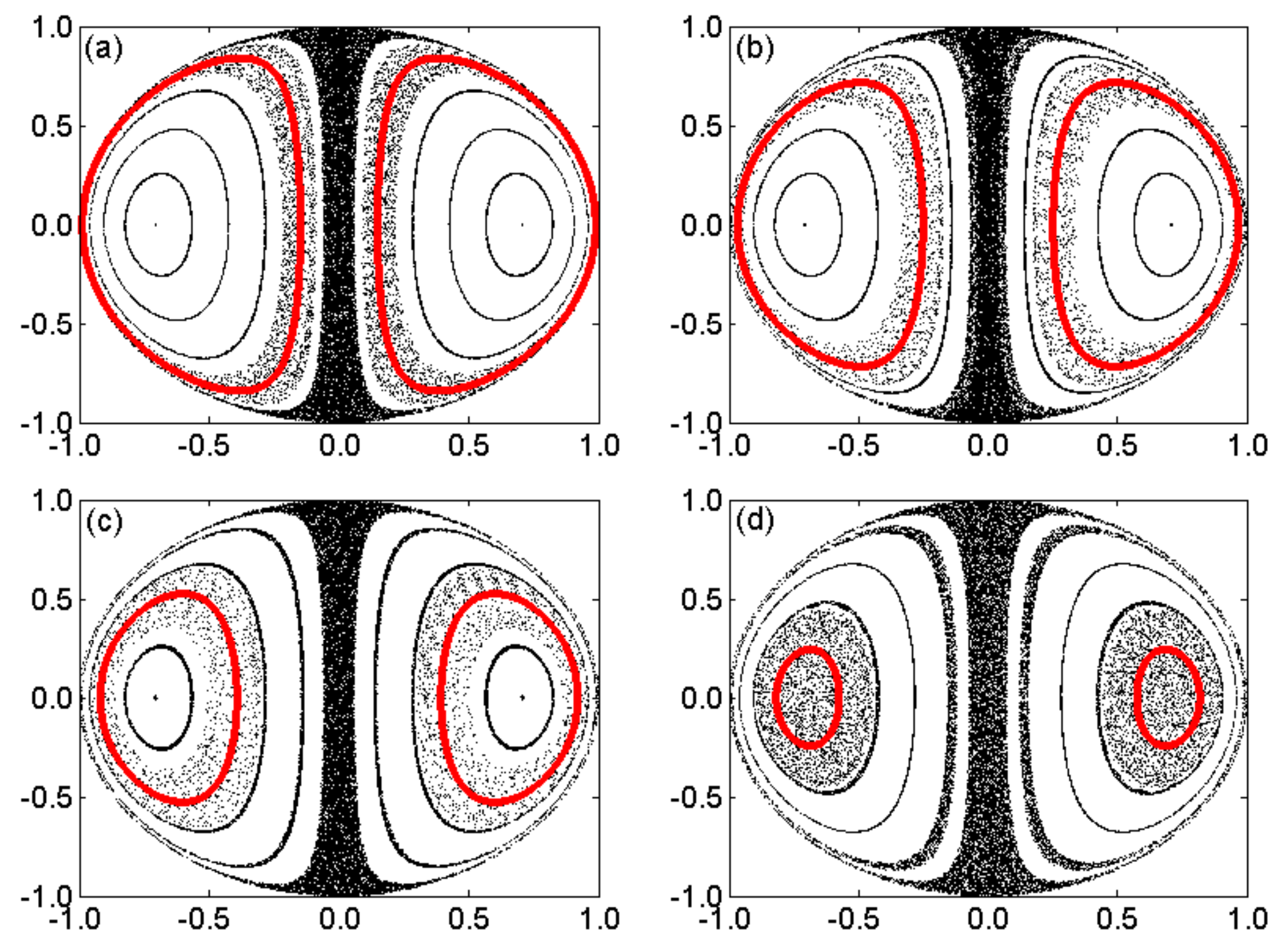}
\end{center}
\caption{\label{figure4} (color online)
Liouvillian sections for the time-dependence ${\mathcal T}_{\omega}$ with the
frequencies $\omega= 0.95, 1.10, 1.25, 1.40$ (a-d) and the amplitude $\epsilon
= 0.05$ with $a_{\omega}(t)={\mathcal T}_{\omega}(t)$. The (red) dashed line
inside the primary mixing region is the torus $\Gamma_{\psi_{1,1}}$.}
\end{figure}

In this paper, we study the influence on mixing for two different
 time-dependent functions of the periodic forcing:
 ${\mathcal C}_{\omega}(t)$ that corresponds to
only one harmonic and ${\mathcal T}_{\omega}(t)$ that is a
triangular shaped forcing (see Fig.~\ref{cos_tri} and
Eqs.~(\ref{cos},\ref{tri})). The two functions differ by:
\begin{equation}
\left| {\mathcal T}_{\omega}\left(t\right)-{\mathcal C}_{\omega}\left(t\right)\right|\leq 0.11, ~~~ \forall t \in \mathbb{R}.
\nonumber
\end{equation}
Figure~\ref{figure2} shows the case where the periodic forcing
$a_{\omega}(t)$ of the rigid body rotation has only  one harmonic
$a_{\omega}(t)={\mathcal C}_{\omega}(t)$.
Here, one can observe two major three-dimensional chaotic mixing regions, all others being
of almost negligible size.  The first one of these major mixing regions is around the pole-to-pole
connections in the center and near the surface of the
droplet. This region corresponds to the resonance of the family of
unperturbed tori $\left\{ \Gamma_{\psi_{n,1}}\right\}_{n}$ with
$n>1$ and is labeled as the central chaotic mixing region. The other
one corresponds to the resonance of the unperturbed tori
$\Gamma_{\psi_{1,1}}$ and is labeled as the primary chaotic mixing
region. While the central mixing region is always present at this
particular location inside the drop, the primary one can be  displaced within
the droplet. For small values of $\omega$, the primary and central
chaotic mixing regions coincide, located around the pole-to-pole
connection. For larger values
of $\omega$, the primary chaotic mixing region separates from the
central one, and penetrates deeper into the droplet.
As $\omega$ is increased further, it moves toward the location of
the circle of degenerate elliptic fixed point of the unperturbed
system ($z=0,~~x^2 + y^2=1/2$) by following the location given by $\psi_{1,1}$.\\
Figure~\ref{figure4} shows the case where the periodic forcing of
the rigid body rotation is $a_{\omega}(t)={\mathcal T}_{\omega}(t)$.
As in the situation where only one harmonic is present, one observes two major
chaotic mixing regions. While the central region remains around the pole-to-pole
connection, the primary one can be shifted across the droplet by varying the
frequency $\omega$, as it is the case for ${\mathcal C}_{\omega}(t)$. \\
Although the triangular amplitude ${\mathcal T}_{\omega}$ is
composed by an infinite number of harmonics, only the first harmonic
plays a non-negligible role.  All other chaotic mixing zones created by higher harmonic
appear of negligible size within the values of parameters studied.
This result reinforces the robustness of the proposed method and confirms that the main parameter influencing the
location of the primary mixing zone is the frequency  $\omega$, and not the
specific shape of the forcing.

\subsection{Control of the size \label{size}}
\begin{figure}
\begin{center}
\includegraphics*[width=14cm]{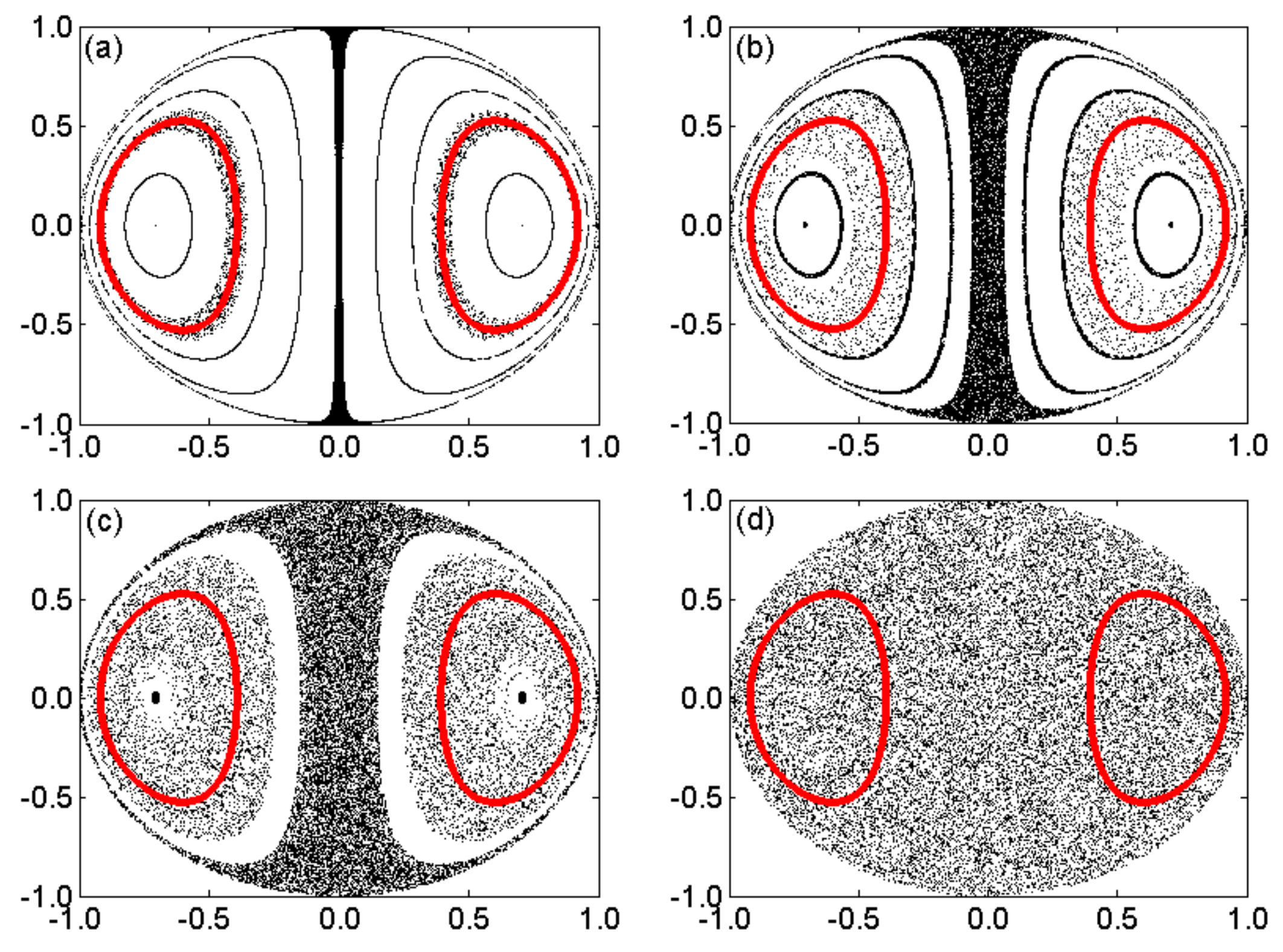}
\end{center}
\caption{\label{figure3} (Color on line)
Liouvillian sections for the time-dependence ${\mathcal C}_{\omega}$ with the frequencies
$\omega= 1.25$ and the amplitude $\epsilon = 0.01, 0.05, 0.10, 0.20$ (a-d). The (red) dashed line inside the primary
mixing region is  $\Gamma_{\psi_{1,1}}$.}
\end{figure}

\begin{figure}
\begin{center}
\includegraphics*[width=14cm]{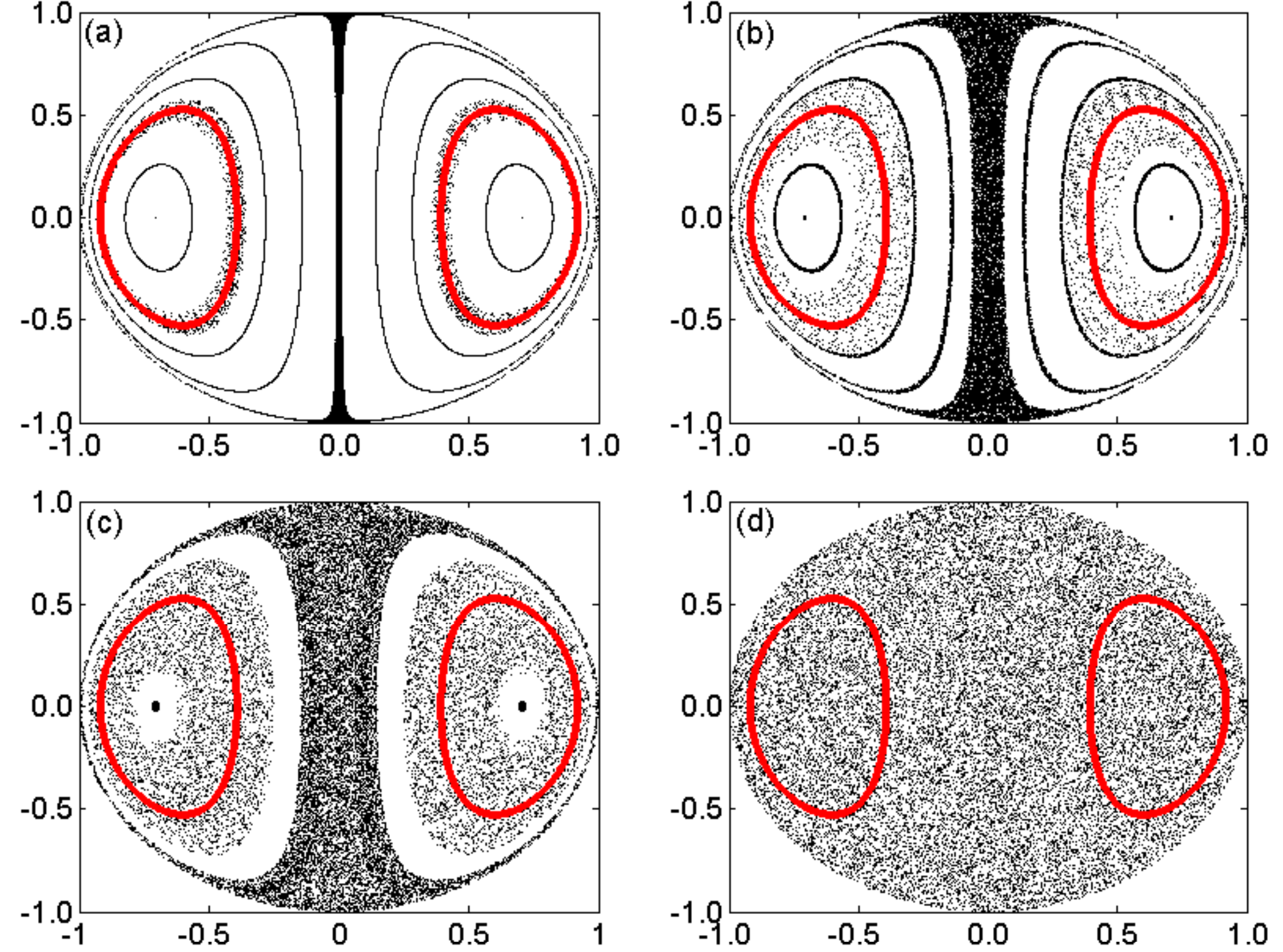}
\end{center}
\caption{ \label{figure5} (Color on line)
Liouvillian sections for the time-dependence ${\mathcal T}_{\omega}$ with the frequencies
$\omega= 1.25$ and the amplitude $\epsilon = 0.01, 0.05, 0.10, 0.20$ (a-d). The (red) dashed line inside
the primary mixing region is  $\Gamma_{\psi_{1,1}}$.}
\end{figure}

Whereas the tuning of the frequency $\omega$ of the rigid body
rotation enables the placement of the primary chaotic mixing region
at a particular location inside the droplet, it is its amplitude
$\epsilon$ that mostly determines the size of the main mixing regions. \\
The sizes of the primary and central chaotic mixing regions are
presented in Figs.~\ref{figure3},\ref{figure5}  for the two
perturbations discussed above.  It is clear that in both cases the
size increases with the amplitude of the rotation $\epsilon$. Such a
control of the size of the mixing allows one to
vary the extent of mixing generated within the droplet
from no mixing to complete mixing.\\
Once again it is important to note that for the case where the periodic
amplitude of rotation is triangular only the first harmonic  plays  a significant role.
Figure~\ref{figure5} shows the case of nearly no mixing ( $\epsilon=0.01$)
and that of complete mixing ($\epsilon=0.20$), the predominance of the first
harmonic is not affected by the magnitude of $\epsilon$.

\section{Conclusions}
In this paper, we have shown that a chaotic mixing region within a droplet is generated by adding a slow oscillatory
rigid-body rotation to the translating drop motion.
Our strategy consists in using resonances between one of the frequencies of
the base flow generated by the steady translation  and the forcing frequency
characteristic of the oscillatory rigid body rotation. This strategy allows for a direct control of both the
location and the size of the mixing by judiciously adjusting the parameters of the
time-dependent rotation, i.e. its amplitude and frequency. One the one hand, controlling the mixing location
could be useful to manufacture inhomogeneous droplets and particles (such as Janus particles) whose properties
could be superior to homogeneous ones.  On the other hand, controlling the size of the mixing zone
could be important to control reaction rates.
Note that in this system, the total mixing state does not have islands of non-mixing zones, as is often
the case with other mixing strategies. Size and location control was obtained not only with a
sinusoidal rotation (with only one harmonic) but also with a rotation consisting of a time dependent triangular function,
with an infinite number of harmonics.  In the latter case, the first harmonic only was shown to play a significant role.

\section*{Acknowledgements}
This article is based upon work partially supported by the NSF
(grants CTS-0626070 (N.A.), CTS-0626123 (P.S.) and 0400370 (D.V.)).
D.V. is grateful to the RBRF (grant 06-01-00117) and to the Donors
of the ACS Petroleum Research Fund. C.C. acknowledges support from
Euratom-CEA (contract EUR~344-88-1~FUA~F) and CNRS (PICS program).


\begin{thebibliography}{}

\end{thebibliography}


\begin{thebibliography}{99}

\bibitem{Ismagilov:2003} H. Song, J.D. Tice and R. F. Ismagilov, A Microfluidic System for Controlling Reaction Networks in Time,
 Angew. Chem. Int. Edit. 42, (2003) 768-772.

\bibitem{Oddy:2001} M.H. Oddy, J.G. Santiago, J.C. Mikkelsen, Electrokinetic Instability Micromixing,
Anal. Chem. 73, (2001) 5822-5832.

\bibitem{Bau:2001} H.H. Bau, J. Zhong and M. Yi, A minute magneto hydrodynamic (MHD) mixer,
Sens. Actuators B 79, (2001) 207-215.

\bibitem{Ouldelmoctar:2003} A. Ould El Moctar, N. Aubry and J. Batton, Electro-hydrodynamic micro-fluidic mixer,
Lab Chip 3, (2003) 273-280.

\bibitem{GlasgowAubry:2003} I.K. Glasgow and N. Aubry, Enhancement of microfluidic mixing using time pulsing,
Lab Chip 3, (2003) 114-120.

\bibitem{Glasgow:2004} I.K. Glasgow, J. Batton and N. Aubry, Electroosmotic mixing in microchannels,
Lab Chip 4, (2004) 558-562.

\bibitem{Goullet:2006} A. Goullet, I.K. Glasgow and N. Aubry, Effects of microchannel geometry on pulsed flow mixing,
Mech. Res. Commun. 33, (2006) 739-746.

\bibitem{Niu:2003} X. Niu and Y-K. Lee, Efficient spatial-temporal chaotic mixing in microchannels,
J. Micromech. Microeng. 13, (2003) 454-462.

\bibitem{Bottausci:2004} F. Bottausci {\it et~al.}, Mixing in the shear superposition micromixer: three-dimensional analysis,
Phil. Trans. Royal Soc. A 362, (2004) 1001-1018.

\bibitem{Stremler:2004} M.A. Stremler, F.R. Haselton and H. Aref, Designing for chaos: applications of chaotic advection at the microscale,
Phil. Trans. Royal Soc. A 362, (2004) 1019-1036.

\bibitem{Lee:2007} Y.K. Lee, C. Shih, P. Tabeling, C.M. Ho, Experimental study and nonlinear dynamic analysis of time-perioidc micro chaotic
mixers,  J. Fluid Mech. 575, (2007) 425-448.

\bibitem{Baj} K. Bajer and H.K. Moffatt, On a class of steady confined Stokes flows with chaotic streamlines,
J. Fluid Mech. 212, (1990) 337-363.

\bibitem{KroujilineandStone:1999} D. Kroujiline and H.A. Stone, Chaotic streamlines in steady bounded three-dimensional Stokes flows,
Physica D 130, (1999) 105-132.

\bibitem{Lee:2000} S.M. Lee, D.J. Im and I.S. Kang, Circulating flows inside a drop under time-periodic nonuniform electric fields,
Phys. Fluids 12, (2000) 1899-1910.

\bibitem{WardandHomsy:2001} T. Ward and G.M. Homsy, Electrohydrodynamically driven chaotic mixing in a. translating drop,
Phys. Fluids 13, (2001) 3521-3525.

\bibitem{Grigoriev:2005} R.O. Grigoriev, Chaotic mixing in thermocapillary-driven microdroplets,
Phys. Fluids 17, (2005) 033601.1-033601.8.

\bibitem{Homsy:2007} X.M. Xu and G.M. Homsy, Three-dimensional chaotic mixing inside drops driven by a transient electric field,
Phys. Fluids 19, (2007) 013102.1-013102.11.

\bibitem{VWG:2007} D. Vainchtein, J. Widloski and R. Grigoriev, Resonant chaotic mixing in a cellular flow,
Phys. Rev. Lett. 99, (2007) 094501.1-094501.4.

\bibitem{WardandHomsy:2003} T. Ward and G.M. Homsy, Electrohydrodynamically driven chaotic mixing in a translating drop part II: Experiments,
Phys. Fluids 15, (2003) 2987-2994.

\bibitem{GSS:2006} R.O. Grigoriev, M.F. Schatz and V. Sharma, Optically controlled mixing in microdroplets,
Lab Chip 6, (2006) 1369-1372.

\bibitem{ChabreyriePRE08} R. Chabreyrie {\it et~al.}, Tailored mixing inside a translating droplet,
Phys. Rev. E 77, (2008) 036314.1-036314.4.

\bibitem{FKP:1988} M. Feingold, L.P. Kadanoff and O. Piro, Passive scalars, three-dimensional volume-preserving maps, and chaos,
J. Stat. Phys. 50, (1988) 529-565.

\bibitem{VVN:1996a} D. Vainchtein, A. Vasiliev and A. Neishtadt, Adiabatic chaos in a two-dimensional mapping,
Chaos 6, (1996) 514-518.

\bibitem{VNM:2006} D. Vainchtein, A. Neishtadt and I. Mezi\'c, On passage through resonances in volume-preserving systems,
Chaos 16, (2006) 043123.1-043123.11.


\bibitem{Lima:1990} R. Lima and M. Pettini, Suppression of chaos by resonant parametric perturbations,
Phys. Rev. A 41, (1990) 726-733.

\bibitem{CFP2:1996} J.H.E. Cartwright, M. Feingold and O. Piro,Chaotic advection in three-dimensional unsteady incompressible laminar flow,
J. Fluid Mech. 316, (1996) 259-284.

\bibitem{AubrySingh:2006} N. Aubry and P. Singh, Influence of particle-particle interactions and particle rotational motions in traveling wave dielectrophoresis,
Electrophoresis 27, (2006) 703-715.

\bibitem{Arn:1988} W.M. Arnold, U. Zimmermann, Electro-rotation: development of a technique for dielectric
 measurements on individual cells and particles, J. electrostat. 21, (1988) 151-191.

\end{thebibliography}
\end{document}